\documentclass[nofootinbib]{revtex4}
\usepackage{amsthm}
\usepackage{amsmath}
\usepackage{graphicx}
\begin{document}
\title{The Necessity of Quantizing Gravity}
\author{Jeremy E. Adelman\\Department of Physics, University of California at Davis \\ One Shields Avenue, Davis, CA, 95616}
\maketitle
\section{The Eppley-Hannah Thought Experiment}
Almost since the beginnings of quantum theory, physicists have endeavored to quantize gravity. These efforts have been remarkably fruitful in yielding novel mathematics and physics, but as yet, no complete and consistent quantum theory of gravity exists \cite{car}. Furthermore, the motivations for producing such a theory remain vague and philosophical,\footnote{See \cite{kief}, page 3-4, for a summary of these motivations} and, of course, there exists at this moment no experimental evidence supporting quantized gravity. Thus, in a sense, a fundamental question remains unanswered: must gravity be quantized? \\
\indent According to a thought experiment proposed by Eppley and Hannah in 1977 \cite{ephan}, the answer is yes. Following in the footsteps of the ``Heisenberg microscope" argument used by Heisenberg as an illustration of the uncertainty principle, Eppley and Hannah consider a particle with momentum uncertainty $\Delta p_i$ and position uncertainty $\Delta x_i$ prepared in the normal minimal uncertainty state
\begin{equation}
\Delta p_i \Delta x_i = \frac{\hbar}{2} 
\end{equation}
If gravity is purely classical, then a gravitational wavepacket of arbitrarily short wavelength $\lambda$ and arbitrarily small momentum $p$ may be prepared and scattered off of the particle. The resulting scattered wavepacket may be then detected and (in principle) used to locate the particle to an uncertainty \begin{equation}
\Delta x_f \approx \lambda
\end{equation} 
in position, while the momentum uncertainty of the particle is no worse than 
\begin{equation}
\Delta p_f \leq \Delta p_i + p
\end{equation}
provided that momentum is conserved. But if $\lambda$ and $p$ are arbitrarily small, then this means
\begin{equation}
\Delta p_f \Delta x_f < \frac{\hbar}{2}
\end{equation}
in violation of the uncertainty principle! Thus, either gravity must be quantized (and hence no longer may a wavepacket have arbitrarily small momentum $p$ for wavelength $\lambda$), or else either the uncertainty principle fails or momentum is not strictly conserved. \\ 
\indent The Eppley-Hannah thought experiment remains controversial for several reasons. First, the entire argument is premised on the requirement the classical gravitational wave collapse the target particle's wavefunction. In their original paper, Eppley and Hannah contend that gravitational radiation that does not collapse the wavefunction would allow for superluminal communication, but this conclusion is disputed (see \cite{kief}, page 16). Furthermore, others have argued (see, for instance, \cite{matt}) that device necessary to conduct the experiment Eppley and Hannah propose is either impossible to construct even in principle, or else incapable of making the measurement Eppley and Hannah require. In this paper, we will endeavor to show both that gravity must indeed collapse a quantum particle's wavefunction and that a modified version of the Eppley-Hannah detector could, at least in principle, be used to yield the contradiction in the original Eppley-Hannah thought experiment.    

\section{Does Gravity Necessarily Collapse the Wavefunction?}
\subsection{The Eppley-Hannah Argument}
In their original paper, Eppley and Hannah made an argument similar to the following for why classical gravitational radiation must necessarily collapse the wavefunction of a quantum particle. Assume to the contrary that gravitational radiation does not collapse the wavefunction. Then, if one prepared two boxes such that a massive particle had an equal likelihood of being in either box and then separates them, a purely gravitational measurement of one of the two boxes would ``detect," necessarily, the superposition state
\begin{equation} \label{state1}
\left|\Psi\right> = \frac{1}{\sqrt{2}} \left(\left|\text{Particle in Box 1} \right> + \left|\text{Particle in Box 2} \right>\right)
\end{equation}
Only after another measurement is taken (say, for instance, by opening the box and peering inside) is the wavefunction then collapsed. For instance, say Box 1 was opened, revealing no particle. Then, a subsequent gravitational measurement of either box would detect the following state
\begin{equation}
\left|\Psi\right> =   \left|\text{Particle in Box 2} \right>
\end{equation}
Eppley and Hannah noted that this change in what the gravitational measurement detects can be used to send messages instantaneously. To illustrate how this would work, assume that the United Federation of Planets wants to launch two simultaneous attacks on two outposts of the Klingon Empire, one that is nearby a Federation starbase, and another that is ten lightyears away.\footnote{Since the original Eppley-Hannah paper precedes the airing of \textit{Star Trek: The Next Generation}, we may safely assume for the purposes of this paper that the Federation and the Klingon Empire are adversaries.} Two squadrons will handle the assault: squadron A will attack the nearer outpost, and squadron B the one that is farther away. However, squadron A's starships have been damaged in previous battles with the Klingon empire, and Starfleet does not know when they will be ready to attack when they dispatch squadron B to prepare for their assault on the other Klingon outpost. If superluminal signals are forbidden, then the Federation will be unable to tell squadron B that squadron A is ready to begin their simultaneous assault without waiting ten years for the signal to propagate. However, suppose two boxes were prepared as above, and one given to each squadron before squadron B departs. The admiral in command of squadron B then monitors his box with a gravitational detector. Whenever the repairs are completed, squadron A's admiral opens his box. Instantaneously, this collapses the wavefunction of both boxes, and no matter whether the particle is revealed to be in A's box or B's box, the gravitational signal admiral B is detecting changes, telling him it is time to begin his assault on the Klingon outpost. \\
\indent There is, however, a problem with this argument, as pointed out by Albers and others (see \cite{Alber}); it does not take into account the true nature of the entangled quantum state. Before admiral A opens his box, the state is  
\begin{equation}
\left|\Psi\right> = \frac{1}{\sqrt{2}} \left(\left|\text{Particle in Box A} \right> + \left|\text{Particle in Box B} \right>\right)\left|\text{Box A Closed} \right>
\end{equation}
When Admiral A opens his box, the state becomes
\begin{equation}
\left|\Psi\right> = \frac{1}{\sqrt{2}} \left(\left|\text{Particle in Box A} \right>\left|\text{A saw Particle} \right> + \left|\text{Particle in Box B} \right>\left|\text{A saw Nothing}\right>\right)
\end{equation}
which, when admiral B takes a gravitational measurement, appears no different from the the previous state. That is to say, no superluminal communication is possible; Admiral B only knows that squadron A has launched its attack after the boxes are brought back together again, but this can only happen subluminally, at which point in time the Klingon outpost squadron B is targeting will have learned of the attack on the other outpost and will have readied its defenses accordingly. \\
\subsection{An Invalid Quantum Operator?}
\indent Since the Eppley-Hannah argument is incumbent on classical gravity collapsing the quantum wavefunction, in order to resurrect it we must make a new argument for its necessity. One such argument is to observe that a classical gravitational detector that does not collapse the wavefunction must itself correspond to a nonlinear operator. To illustrate this, let us presume that the gravitational detector used by Admiral B in the above example returns the mass of the box, and define the measurement as corresponding to operator $\hat{O}$. Thus, for $\left|\Psi\right>$ as defined in equation \ref{state1}, the action of $\hat{O}$ is
\begin{equation}
\hat{O}\left|\Psi\right> = \frac{m}{2}\left|\Psi\right>
\end{equation}
where $m$ is the mass of the particle. But as per equation \ref{state1}
\begin{equation}
\left|\Psi\right> = \frac{1}{\sqrt{2}} \left(\left|\text{Particle in Box A} \right> + \left|\text{Particle in Box B} \right>\right)
\end{equation}
meaning that if $\hat{O}$ is linear
\begin{equation} 
\begin{split}
\hat{O}\left|\Psi\right> &= \frac{1}{\sqrt{2}} \left(\hat{O}\left|\text{Particle in Box A} \right> + \hat{O}\left|\text{Particle in Box B} \right>\right) \\ &=  \frac{1}{\sqrt{2}} \left(0\left|\text{Particle in Box A} \right> + m\left|\text{Particle in Box B} \right>\right) \\ &= \frac{m}{\sqrt{2}}\left|\text{Particle in Box B}\right> \\ &\neq \frac{m}{2}\left|\Psi\right> = \hat{O}\left|\Psi\right>
\end{split} 
\end{equation}
a contradiction. Thus, since operator $\hat{O}$ is nonlinear, it cannot correspond to a measurement, as per the standard definition of a measurement in quantum mechanics. Notice too that the precise definition of how the operator $\hat{O}$ is defined is irrelevant; the assumption that gravity does not collapse the wavefunction implies that every state $\left| \Psi \right>$ must be an eigenstate of any operator corresponding to a gravitational-based measurement (because otherwise, the gravitational ``measurement" would change the state). However, the only such operators that are linear are those proportional to the identity - that is to say, measurement where all states return the same eigenvalue, and thus where no information is gleaned by taking the measurement. \\
\indent That such a gravitational measurement is not permitted given the postulates of standard quantum mechanics is not, however, necessarily damning, since a quantum system with classical gravity should, by its very nature, be an extension of normal quantum mechanics. Nonetheless, it should be pointed out that, in order to accommodate measurement with gravity that do not collapse the wavefunction, nonlinear (and thus non-Hermitian) operators corresponding to observables must be allowed. 
\subsection{A Modified Eppley-Hannah Argument for the Necessity of Gravitational Collapse of the Wavefunction}
\indent In order to prove that a gravitation measurement must necessarily collapse the wavefunction, a new argument is needed. Let us again consider two boxes, prepared as before, such that a particle of mass $m$ has an equal probability of being in either box $A$ or box $B$ when the boxes are separated. We then take box $B$ and throw it into a Schwarzschild black hole of mass $M$. Notice that the black hole's response to ``eating" the box constitutes a purely gravitational measurement of the box's mass; by measuring the change in the black hole's mass (for instance by measuring the peak frequency of the Hawking radiation it emits) we can determine the mass of the box that was tossed inside. If gravity collapses the wavefunction, then we will observe that the black hole's mass increases either by the mass of the box itself or by the mass of the box plus $m$, depending on whether the particle was inside the box or not (we can, of course, then confirm this measurement by opening box $A$). \\
\indent Assume to the contrary, however, that gravity does not collapse the wavefunction. In that case, the mass of the black hole necessarily increases by the mass of the box plus $\frac{m}{2}$ when we measure it. Now let us open box $A$, a measurement that necessarily collapses the wavefunction; that is to say, we either observe that the particle is inside box $A$ or it is not. In either case, however, observe that, in order for the universe to conserve mass-energy, the observed mass of the black hole \textit{must change} after we open box $A$, either by decreasing by $\frac{m}{2}$ if we see the particle or by increasing by $\frac{m}{2}$ if we do not. \\
\indent Of course, the time at which the discontinuity in the black hole mass occurs poses something of a problem; should the change be ``instantaneous" in some reference frame, then that would imply superluminal communication, in exactly the same manner as the original Eppley-Hannah argument. However, even if the discontinuity is measured after light has had a chance to travel from the opened box $A$ to the black hole that ate box $B$, there is still a problem. Let us consider the situation where the particle was observed to be in box $A$ (if this was not what we observed, we could rerun the experiment as many times as necessary until it was so observed). In that case, the mass of the black hole is observed to shrink by $\frac{m}{2}$. But this, of course, means that the surface area, and thus the entropy, of the black hole decreases by an amount proportional to $mM$, where $M$ is the mass of the black hole itself. Meanwhile, the act of opening box $A$ and seeing whether the particle is inside corresponds to an increase of informational entropy of one bit. Thus it is clear that by judiciously choosing $m$ and $M$, we can construct a situation in which finding the particle in box $A$ decreases the entropy of the universe, in violation of the second law of thermodynamics. \\
\indent Thus, we conclude that if gravity does not collapse the wavefunction, then either the universe does not conserve mass-energy or the second law of thermodynamics is violated. Assuming both the conservation of mass-energy and the second law thus implies that gravity necessarily collapses quantum wavefunctions, the necessary condition for the Eppley-Hannah argument.   
\section{Problems with the Eppley-Hannah Detector}
\subsection{The Meaning of ``Detection"}
A question not addressed in either Eppley-Hannah's paper or the original Heisenberg microscope from which their argument is derived is what, precisely, comprises a detection of the low energy classical wave. To remedy this deficiency, we will adopt the following definition for a detection:
\newtheorem{newdef}{Definition}
\begin{newdef} \label{detdef}
A signal is \textbf{detected} by a device if the resulting device output may be amplified arbitrarily.
\end{newdef}
\noindent Using this definition, both thought experiments fail to actually detect the classical wave signal. In Heisenberg's original argument (see \cite{Heisen}, page 21), the classical EM wave is merely \textit{resolved} to below the uncertainty principle limit by a lens; there is no device there to detect the wave once it is so resolved, and no argument made that such a resolved wave could be detected as per definition \ref{detdef}. In the Eppley-Hannah experiment, the classical gravitational wave is ``detected" by an array of quantum harmonic oscillators, whose masses and spring constants are sufficiently large and small respectively such that the energy level spacing is on order the energy of the gravitational wavepacket. Eppley and Hannah argue that one of these harmonic oscillators undergoing a transition out of its ground state constitutes a ``detectable transition," but in what sense (i.e., by what other detector; say, the human eye) is never specified; we contend that because the Eppley-Hannah detector does not \textit{amplify} the signal, the proposed Eppley-Hannah detector does not qualify as actually making a detection! 
\subsection{A Black Hole?}
A second problem with the Eppley-Hannah detector was noted by Mattingly in 2006 \cite{matt}: the detector Eppley and Hannah propose in their paper has a radius smaller than its own Schwarzschild radius! Thus the Eppley-Hannah detector is a black hole, hence even if the detector could make a detection that localizes a target particle in a way that violates the uncertainty principle, there would be no way of extracting this information. Obviously, this poses a problem for the Eppley-Hannah thought experiment; any hopes of resurrecting the experiment will need to both propose a new detector capable of detecting the gravitational wavepackets and demonstrate that the proposed detector is not itself a black hole.  
\section{The Modified Eppley-Hannah Experiment}
\subsection{Overview}
In order to resurrect the Eppley-Hannah experiment, we make several modifications:
\begin{enumerate}
\item Instead of an array of detectors as in the Eppley-Hannah thought experiment, our experimental setup will consist of a single detector. This will allow us to localize the experiment to a much smaller region of space, allowing us to avoid the $O\left(\frac{1}{R^2}\right)$ suppression of the gravitational wavepacket energy density at distance $R$ and also the temptation to pile multiple detectors on top of each other until they form a black hole. The cost of this decision is that we now must run our experiment a large number of times (or, equivalently, build and run a large number of spacelike separated experiments) in order to achieve a positive detection.
\item Rather than a harmonic oscillator, our detector will consist of a particle (possible composite) of mass $m$ and charge $q$ bound in a finite spherical, approximately square well electrostatic potential. Notice that this means our detector amplifies the gravitational wave signal by producing (on detection) a massive particle with a larger energy then that of the incident gravitational wavepacket. 
\item Localizing the particle is then reliant on using time of flight data for the classical gravitational wave packet that is scattered off of the target particle. Notice that this now requires us to have synchronized, arbitrarily accurate clocks, a requirement that comes with its own set of issues that will not be discussed in this paper.
\end{enumerate} 
\subsection{Experiment Design}
\begin{figure}
\includegraphics[width=\linewidth]{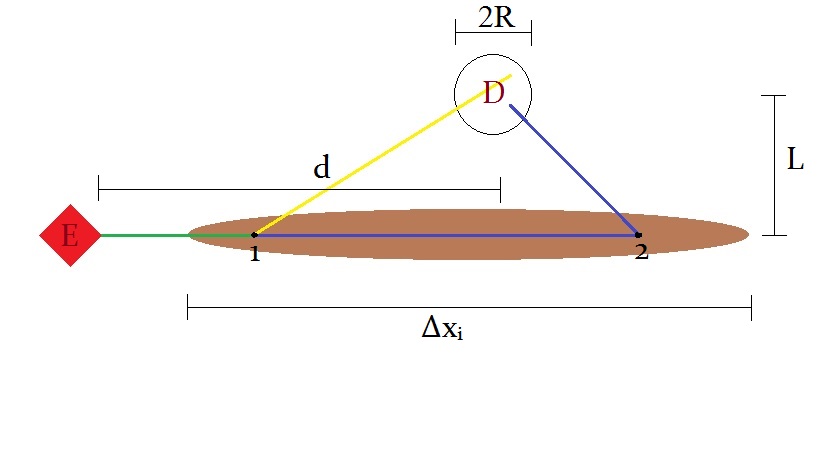}
\caption{The modified Eppley-Hannah experiment. The blue and yellow paths show two possible trajectories of the gravitational wave packet depending on where it encounters and scatters off of the target particle.}
\label{traj}
\end{figure}

Our modified Eppley-Hannah experimental setup is illustrated in figure \ref{traj}. A gravitational wavepacket of arbitrarily small momentum $p$, wavelength $\lambda$, and spatial extent (presumably $\approx \lambda$ in all three spatial dimensions) is emitted from the emitter (E), scatters off a target particle with spatial uncertainty $\Delta x_i$ in the direction the packet is initially traveling, and then is detected by the detector (D) located a distance $d$ away from the emitter along the emitted wave axis and a distance $L$ away from the axis. Whether the particle is at point 1 or point 2 can be distinguished by comparing the time of flight between emission and detection as measured by synchronized clocks at the two locations, since the trajectories (yellow and blue) are different lengths.
\subsection{The Modified Detector}
\subsubsection{Detector Design}
\begin{figure}
\includegraphics[width=\linewidth]{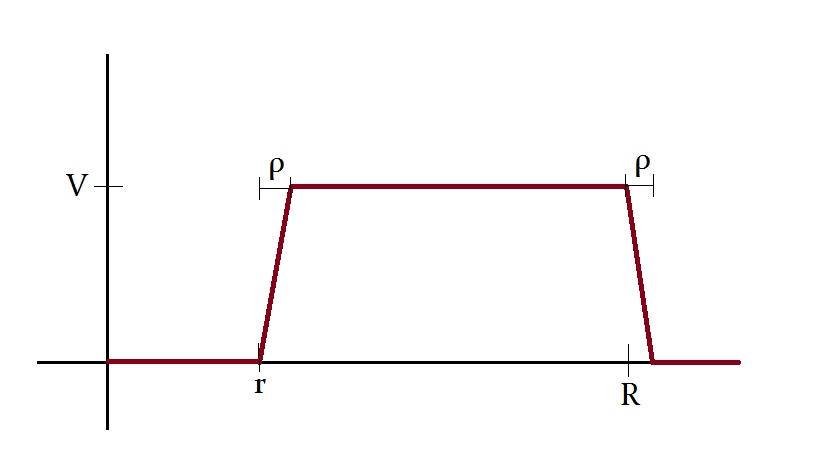}
\caption{The detector potential energy profile as a function of radial distance from the center. Notice that this is not drawn to scale; for the real detector $\rho << r << R$.}
\label{detector}
\end{figure}

The detector itself consists of a massive, charged particle bound in a spherically symmetric potential with a radial profile as given in  figure \ref{detector}.
In this experiment, it is presumed that 
\begin{equation}
\rho << r << R
\end{equation}
That is to say, the local bound state of the massive particle is approximately that of a spherical square well. The spherical finite well is a well known problem from basic quantum mechanics and has the solution of spherical Bessel functions inside the well and spherical Hankel functions in the classically forbidden region. We will further configure the well such that there is exactly one bound state for the massive particle, a state whose energy $E$ is very close to $V$, the potential energy of the well barrier; this is done so that the energy necessary to trigger a detection $\epsilon = V - E$ is as small as possible. From the boundary conditions on the radial square-well solution, the ground state energy must satisfy the equation
\begin{equation}
-\cot\left(\sqrt{\frac{2mE}{\hbar^2}}r\right)=\sqrt{\frac{\epsilon}{E}}
\end{equation} 
which implies, if $\epsilon << E$ as required, that
\begin{equation} \label{rsize}
r = \frac{\hbar}{\sqrt{2mE}}\left(\frac{\pi}{2}+\sqrt{\frac{\epsilon}{E}} + O\left(\frac{\epsilon}{E}\right)^\frac{3}{2} \right) = \frac{\hbar}{\sqrt{2mV}}\left(\frac{\pi}{2}+\sqrt{\frac{\epsilon}{V}} + O\left(\frac{\epsilon}{V}\right) \right)
\end{equation}\\ 
In order for a detection to occur, we must have the classical wavepacket scatter towards the single detector and deposit an energy of $\epsilon$ to excite the bound particle into its free excited state. Both are in all likelihood highly improbable events that depend on the sundry experimental parameters $\epsilon$, $L$, etc. For the purposes of this paper, though, it is sufficient for us to assume that this probability $P$ is nonzero provided the gravitational wavepacket has sufficient energy to trigger the transition to the excited state.  
\subsubsection{``No Black Hole" Restrictions on Detector Parameters}
To avoid the issue raised by Mattingly vis-\`{a}-vis the original Eppley-Hannah detector, we must now put restrictions in place to ensure that our detector is not a black hole. The first restriction is on the mass of the bound particle. The bound particle occupies a radial distance roughly equal to the classically allowed region $r$ plus the attenuation distance in the classically forbidden region. In the classically forbidden region, the massive particle's wavefunction falls off like the Hankel function
\begin{equation}
\left|\Psi\left(x_r\right)\right|^2 \propto \frac{\hbar^2}{2m\epsilon x_r^2}e^{-\frac{2\sqrt{2m\epsilon}x_r}{\hbar}}
\end{equation}
or, after multiplying by the $x_r^2$ that we would pick up in the spatial integral, we get that the probability of the massive particle being a distance $x_r$ into the forbidden region goes as
\begin{equation}
P_{forbidden} \propto e^{-\frac{2\sqrt{2m\epsilon}x_r}{\hbar}}
\end{equation}
meaning that the particle is localized to a radial distance roughly on order
\begin{equation}\label{atten}
\delta_{x_r} = \frac{\hbar}{\sqrt{2m\epsilon}} + r
\end{equation}
Thus we have that
\begin{equation} \label{blackhole1}
\frac{\hbar}{\sqrt{2m\epsilon}} + r > r_s = \frac{2Gm}{c^2}
\end{equation}
where $r_s$ is the usual Schwarzschild radius of a particle of mass $m$. Notice that equation \ref{rsize} and the assumption $V >> \epsilon$ means that the restriction here is functionally
\begin{equation}
\frac{\hbar}{\sqrt{2m\epsilon}} > \frac{2Gm}{c^2}
\end{equation} or
\begin{equation}\label{bholefunct}
\epsilon < \frac{\hbar^2c^4}{8G^2m^3}
\end{equation} \indent However, this is not the only source of energy density; the potential energy profile in figure \ref{detector} can also yield a black hole if restrictions are not applied. Presuming this potential energy profile is produced electrostatically, this requires a potential
\begin{equation}
\Phi\left(x_r\right)= \begin{cases} 0 &\text{if $x_r < r$ or $x_r > R+\rho $} \\ \frac{V}{q\rho}x_r -\frac{Vr}{q\rho}  &\text{if $r \leq x_r \leq r+\rho $} \\ \frac{V}{q} &\text{if $r+\rho < x_r < R $} \\ -\frac{V}{q\rho}x_r+\frac{V\left(R+\rho\right)}{q\rho} &\text{if $R \leq x_r \leq R+\rho $}\end{cases}
\end{equation}
implying a radial electric field
\begin{equation}
E_r\left(x_r\right) = \begin{cases} \frac{V}{q\rho} &\text{if $r < x_r < r +\rho $} \\ -\frac{V}{q\rho} &\text{if $R < x_r < R +\rho $} \\ 0 &\text{otherwise} \end{cases}
\end{equation}
This gives an energy density of
\begin{equation}
U\left(x_r\right) = \begin{cases} \frac{1}{8\pi}\left(\frac{V}{q\rho}\right)^2 &\text{if $r < x_r < r +\rho $ or $R < x_r < R +\rho $} \\ 0 &\text{otherwise} \end{cases}
\end{equation}
or a total energy in the full detector of
\begin{equation}
 \begin{split} E_{det} =& 4\pi\left(\int_{r}^{r+\rho}\frac{V^2x_r^2}{8 \pi q^2\rho^2} dx_r+ \int_{R}^{R+\rho}\frac{V^2x_r^2}{8 \pi q^2\rho^2} dx_r\right) \\ =& \frac{V^2}{6q^2\rho^2}\left(3R^2\rho +3R\rho^2 +3r^2\rho + 3r \rho^2 + 2\rho^3\right)
\end{split} \end{equation}
Thus, the ``no black hole" restriction on the whole detector requires
\begin{equation} 
R > \frac{GV^2}{3c^4q^2\rho^2}\left(3R^2\rho +3R\rho^2 +3r^2\rho + 3r \rho^2 + 2\rho^3\right)
\end{equation}
or
\begin{equation}
\left(\frac{q}{V} \right)^2 > \frac{G}{3c^4}\left(3\frac{R}{\rho} +3 +3\frac{r^2}{R\rho} + 3 \frac{r}{R} + 2\frac{\rho}{R}\right) 
\end{equation}
which, since
\begin{equation}
R >> r >> \rho
\end{equation} means this restriction is functionally
\begin{equation}\label{blackhole2}
\left(\frac{q}{V} \right)^2 > \frac{GR}{c^4\rho}
\end{equation}
Limiting ourselves to just the ``inner shell" of the detector (that is, the energy density between $r$ and $r + \rho$, we have an energy of
\begin{equation}
E_{inner} = \frac{V^2}{6q^2\rho^2}\left(3r^2\rho + 3r \rho^2 + \rho^3\right)
\end{equation}
and thus a ``no black hole" restriction of
\begin{equation} 
r > \frac{GV^2}{3c^4q^2\rho^2}\left(3r^2\rho + 3r \rho^2 + \rho^3\right)
\end{equation}
or
\begin{equation}
\left(\frac{q}{V}\right)^{2} > \frac{G}{3c^4}\left(3\frac{r}{\rho} + 3 + \frac{\rho}{r}\right)
\end{equation}
which, again, is functionally
\begin{equation} \label{blackhole3}
\left(\frac{q}{V}\right)^{2} > \frac{Gr}{c^4\rho}
\end{equation}
but of course $R >> r$, so if equation \ref{blackhole2} is satisfied, then so is equation \ref{blackhole3}, meaning it is only the ``outer shell" equation that must concern us.

\subsubsection{Detector Stability}
If left to its own devices, the massive particle trapped in our detector will eventually leave the detector via quantum tunneling. This, of course, would be indistinguishable from a detection of the gravitational wavepacket, and thus we must configure our detector such that the probability of a quantum tunneling event while the experiment is running is orders of magnitude less than the probability of detecting a wavepacket. The tunneling probability of an incident wave of energy $E$ in the bound ($x_r < r$) region of the energy profile in figure \ref{detector} is estimated by the usual method\footnote{Notice that this is actually an overestimate of the tunneling probability, since the actual radial solution is a Hankel function, suppressed by an additional factor of $\frac{\hbar}{\sqrt{2m\epsilon}R}$} to be
\begin{equation}
T = e^{-\frac{2\left(R-r\right)}{\hbar}\sqrt{2m\epsilon}}
\end{equation} 
which means, given a frequency 
\begin{equation}
\omega = \frac{E}{\hbar}
\end{equation}
we have a tunneling probability as a function of time of
\begin{equation}
T\left(t\right) \approx \frac{Et}{\hbar} e^{-\frac{2R}{\hbar}\sqrt{2m\epsilon}}
\end{equation}
To run the experiment, we need the detector to be stable over time of flight difference between a detection at the front of the region the target particle might occupy and the back of said region. From figure \ref{traj}, it is clear that provided the detector is placed in the middle of the $\Delta x_i$ region, this time of flight difference between a front signal and a back signal is 
\begin{equation}
t = \frac{\Delta x_i}{c}
\end{equation}
hence the probability of a false detection is
\begin{equation}
T_{false} \approx \frac{E\Delta x}{\hbar c} e^{-\frac{2R}{\hbar}\sqrt{2m\epsilon}}
\end{equation}
which we need to be less than the probability of an actual signal, namely that
\begin{equation} \label{Nofalse}
\frac{E\Delta x_i}{\hbar c P} e^{-\frac{2R}{\hbar}\sqrt{2m\epsilon}} < 1
\end{equation}
where, as before, $P$ is the probability of a positive detection of the gravitational wavepacket.

\subsection{Running the Experiment}
We will assume that the target is initially prepared in a minimal uncertainty state, implying
\begin{equation}
\Delta p_i = \frac{\hbar}{2\Delta x_i}
\end{equation}
Now let us presume that our detector makes a detection of the scattered gravitational wavepacket. Provided we are in the limit $\lambda << r$ (which since $\lambda$ can be arbitrarily small, there is no reason for us not to be), the detector can only localize the detection to be within the region occupied by the massive, bound particle (both where this particle is classically forbidden and where it is classically allowed).
Therefore, the uncertainty in the time of flight value recorded is
\begin{equation}
\Delta t = \frac{2\hbar}{c \sqrt{2m\epsilon}} + \frac{2r}{c}
\end{equation}
We now observe (see figure \ref{traj}) that the time-of-flight as a function of particle position is \begin{equation}
t = \frac{1}{c}\left[\sqrt{\left(d-x\right)^2+L^2}+x\right]
\end{equation} 
hence
\begin{equation}
\Delta t = \frac{1}{c}\left[1+\frac{x-d}{\sqrt{\left(d-x\right)^2+L^2}}\right]\Delta x
\end{equation}
which implies, since 
\begin{equation}
\frac{x-d}{\sqrt{\left(d-x\right)^2+L^2}} < 1
\end{equation}
that
\begin{equation} \label{dxf}
\Delta x_f < 2c\Delta t = \frac{4\hbar}{\sqrt{2m\epsilon}} + 4r
\end{equation}
Now let us consider the final momentum uncertainty. The incident gravitational wavepacket has momentum $p$. By starting the target particle in a zero momentum state, we ensure that the momentum of the scattered wavepacket is no more than $p$. Setting
\begin{equation}\label{momentum}
p = \frac{a\epsilon}{c}
\end{equation}
for unitless parameter $a > 1$ (that is, setting the initial momentum of the wavepacket to be above the trigger threshold of the detector) then implies that the final momentum uncertainty is
\begin{equation}
\Delta p_f \approx p + \Delta p_i = \frac{a\epsilon}{c} + \Delta p_i
\end{equation} 
 hence we want to show that for a valid choice of detector parameters and $a$
\begin{equation}\label{limit}
\left(\frac{4\hbar}{\sqrt{2m\epsilon}} + 4r\right)\left(\frac{\hbar}{2\Delta x_i} + \frac{a\epsilon}{c} \right) < \frac{\hbar}{2}
\end{equation}
that is to say, the exists a valid configuration for our modified Eppley-Hannah experiment in which we are able to locate the target particle to below the uncertainty principle limit.

\section{Beating the Uncertainty Principle Limit}
Plugging the results for $r$ in equation \ref{rsize} into equation \ref{limit} gives
\begin{equation}
4\left(\frac{\hbar}{\sqrt{2m\epsilon}} + \frac{\pi\hbar}{2\sqrt{2mV}}\left[1 + O\left(\sqrt{\frac{\epsilon}{V}}\right) \right]\right)\left(\frac{\hbar}{2\Delta x_i} + \frac{a\epsilon}{c} \right) < \frac{\hbar}{2}
\end{equation} 
Since $\epsilon << V$, this is functionally
\begin{equation}
\frac{4\hbar}{\sqrt{2m\epsilon}}\left(\frac{\hbar}{2\Delta x_i} + \frac{a\epsilon}{c} \right) < \frac{\hbar}{2}
\end{equation}
If we now start with our target particle in a squeezed state, we may also safely presume that
\begin{equation}
\frac{\hbar}{2\Delta x_i} << \frac{a\epsilon}{c} 
\end{equation} or
\begin{equation} \label{squeeze}
\Delta x_i >> \frac{\hbar c}{2 a \epsilon}
\end{equation}
Our requirement to beat the uncertainty limit then becomes, functionally
\begin{equation}
\epsilon < \frac{mc^2}{32a^2} \label{ep1}
\end{equation} This, coupled with the functional ``no black hole" restriction in equation \ref{bholefunct}
\begin{equation} \label{ep2}
\epsilon < \frac{\hbar^2c^4}{8G^2m^3}
\end{equation}
implies that the uncertainty principle is validly violated in the limit that $\epsilon$ gets very small.
\section{Example Values of the Detector Parameters for a Viable Uncertainty Violating Experiment}
To show that a valid uncertainty-violating experiment is possible to construct (at least in theory), we will now attempt to assign values to the various detector parameters. As our analysis will clearly illustrate, in a situation where there are no restrictions on the values of $r$, $m$, and $V$, an experimental apparatus that violates the uncertainty principle is easily constructed. However, if we wish to impose restrictions on these values, such as requiring that $r$ be at least as large as the atomic scale ($r >  10^{-10} \  \mathrm{m}$), that $m$ be at least as large as the electron mass ($m > 10^{-31} \ \mathrm{kg}$) and requiring $V$ to have the ability to be made large pursuant to definition \ref{detdef}, we find no such detector can be constructed. This result follows directly from equation \ref{rsize}, which requires that
\begin{equation} \label{vprob}
V \approx \frac{\pi^2 \hbar^2}{8mr^2} 
\end{equation}
Clearly, then, to make $V$ arbitrarily large, we must be able to make either $m$ or $r$ arbitrarily small, something that is impossible to do if both are bounded from below. Since we want to preserve our definition of a detection, we must then either throw out our restriction on $m$ or on $r$. Of the two, it is philosophically preferable to be in the limit of very small $m$. Presuming $m << 10^{-31} \ \mathrm{kg}$ merely amounts to assuming that there exists, in addition to the know physical particles of our universe, a new, (sufficiently) stable charged particle with a mass below that of the electron. Logically, the existence or nonexistence of such a particle should have no bearing on whether or not gravity is quantized, meaning that, while our thought experiment could not be run in this universe, it could be run in a universe that differs from ours only in a way that, at least at face value, should have no effect on whether or not gravity must be quantized. \\
\indent Thus, let us restrict $r$ to being atomic scale ($r \approx 10^{-10}\ \mathrm{m}$). In this case, to have $V$ be on the electoweak scale ($V \approx 10^{-13} \mathrm{J}$, where, presumably, we could then increase the size of the signal by detecting the now free particle with another amplifying detector), would require by equation \ref{vprob} that
\begin{equation}
m \approx 10^{-35} \ \mathrm{kg}
\end{equation} 
Clearly in this limit, equation \ref{ep1} is the dominant restriction on $\epsilon$, and thus, for $a \approx 1$,
\begin{equation}
\epsilon \approx 1 \ \mathrm{eV} \approx 10^{-19} \  \mathrm{J}
\end{equation}
would be the order of magnitude for the largest uncertainty violating $\epsilon$. Notice that this is sufficiently small that we are in the limit \begin{equation} 10^{-13} \ \mathrm{J} \approx V >> \epsilon \approx  10^{-19} \ \mathrm{J}  \end{equation} as required. With this as our $\epsilon$, to satisfy equation \ref{squeeze}, we require \begin{equation} 
\Delta x_i >> 10^{-7} \ \mathrm{m}
\end{equation} 
which means even a millimeter scale $\Delta x_i$ will suffice. \\
\indent This leaves only the second ``no black hole" requirement in equation \ref{blackhole2} and the ``no false positive" restriction of equation \ref{Nofalse} left to satisfy. The latter requires that
\begin{equation}
\frac{E\Delta x_i}{\hbar c P} e^{-\frac{2R}{\hbar}\sqrt{2m\epsilon}} < 1
\end{equation}
Clearly, without better knowledge of the value of $P$ (beyond it being very small), it is impossible to ascertain an appropriate value for $R$. However, we do observe that 
\begin{equation}
\frac{2\hbar}{\sqrt{2 m \epsilon}} \approx 10^{-7} \ \mathrm{m}
\end{equation}
meaning that for $R \approx 1 \ \mathrm{m}$, the above is well satisfied for 
\begin{equation}
P > 10^{-4300000}
\end{equation} which is a sufficiently small number that it is reasonable to presume that the ``no false positive" restriction is satisfied for meter scale $R$. The ``no black hole" restriction, which requires that
\begin{equation}
q > \sqrt{\frac{GR}{c^4\rho}}V
\end{equation}
now gives (presuming $\rho \approx 10^{-13} \  \mathrm{m}$, just below the atomic scale) that
\begin{equation}
q > 10^{-37} \ \mathrm{statC}
\end{equation}
which is twenty-eight orders of magnitude smaller than the fundamental charge $e$, meaning $q = e$ is more than sufficient to satisfy the restriction in equation \ref{blackhole2}. \\
\indent To summarize, we have shown that the following choices for the detector parameters give rise to a viable, non-black-hole forming, uncertainty violating Eppley-Hannah like experiment
\begin{equation}
\begin{split}
m &\approx 10^{-35} \ \mathrm{kg} \\
q &= e \\
R &\approx 1 \ \mathrm{m} \\
r &\approx 10^{-10} \ \mathrm{m} \\
\rho &\approx 10^{-13} \ \mathrm{m} \\
\Delta x_i &\approx 10^{-3} \ \mathrm{m} \\
V &\approx E \approx 10^{-13} \ \mathrm{J} \\
\epsilon &\approx 10^{-19} \ \mathrm{J} \\
\end{split}
\end{equation}
subject to the addition requirement that the wavelength of the incident gravitational wavepacket be smaller than the detector scale
\begin{equation}
\lambda << r \approx 10^{-10} \ \mathrm{m}
\end{equation}
\section{Conclusion}
Our results suggests that the Eppley-Hannah thought experiment can be resurrected from the criticisms levied against it by Mattingly and Albers, and that there exists ample parameter space in which we can show that our modified experiment would violate the uncertainty principle. However, our modified thought experiment does suffer from the requirement of the existence of a charged, (reasonably) stable particle with charge $e$ and a mass several orders of magnitude less than that of the electron. Since such a particle does not appear to exist in our universe, our modified thought experiment really only applies to a similar universe differing from ours by the existence of such a particle, the existence or nonexistence of which presumably has no bearing on whether or not gravity must be quantum and not classical. \\
\indent In addition to this obvious criticism, there remain several other obvious lines of attack against our argument that gravity must be quantized (or else either conservation of momentum, conservation of energy, the uncertainty principle, or the second law of thermodynamics must be discarded). First, our argument relied on the existence of a classical gravitational wavepacket of arbitrarily small wavelength and spatial extent. While such a wavepacket is in principle allowed to exist in a universe with classical gravity, perhaps it is possible to show that no physical apparatus could actually construct such a wavepacket. Secondly, our argument has not considered the uncertainties associated with using real synchronized clocks; these, obviously, would change the details of the analysis, however since we can configure our experiment to violate the uncertainty principle not just modestly but by orders of magnitude (by, for instance, making $\epsilon$ much smaller than is necessary), it stands to reason that the additional complications arising from the clocks would not be fatal to the general premise of the argument. Finally, there exists the possibility that $P$ is much smaller than we presumed in this paper. Provided $P$ is non-zero, of course, it is always possible to reconfigure the parameters in such a way the equation \ref{Nofalse} is satisfied, though this would likely come at the expense of our detector having an atomic scale $r$ (by making $r \approx l_p$, it is easy to achieve false detection probabilities of one in $10^{10^{25}}$ and larger). However, it may be possible to show that $P$ is, in fact, identically zero, which would of course negate the entire argument. \\
\indent As pointed out by Callender and Huggett \cite{calhug} however, whether or not one can successfully argue via some sort of Eppley-Hannah-esque thought experiment that gravity \textit{must} be quantized, there still remains strong physical and philosophical motivations suggesting gravity \textit{should} be quantized, more than enough to justify continued research in pursuit of a complete, consistent theory of quantum gravity.
\section{Acknowledgment}
The author acknowledges funding from the Department of Energy, grant number DE-FG02-91ER40674
\bibliographystyle{ieeetr} 
\bibliography{TheNecessityofQuantGrav}

\end{document}